RESEARCH ARTICLE

# Measuring trade costs and analyzing the determinants of trade growth between Cambodia and major trading partners: 1993–2019

Borin Keo *, Bin Li*, Waqas Younis

School of Economics and Trade, Hunan University, Changsha, China

* keo.borin@yahoo.com (BK); libin43@sohu.com (BL)

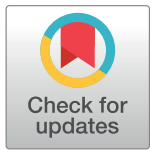

## Abstract

High trade costs pose substantial barriers to the process of trade liberalization. This study aims to measure trade costs and explore the driving forces behind the growth of bilateral trade between Cambodia and its top 30 trading partners from 1993 to 2019. Using a micro-founded measure of trade costs derived from the gravity model, we find that Cambodia's average trade costs decreased by 35.43 percent between 1993 and 2019. Fluctuations in average trade costs persisted until 2014, despite Cambodia's accession to the World Trade Organization (WTO) in 2004. Since then, these costs have declined more rapidly. Cambodia's bilateral trade costs are lower with its major trading partners in Southeast Asia and East Asia than with those in South Asia, Oceania, Europe, and North America. Cambodia's average trade costs with developing and emerging economies are lower than those with developed economies. Between 2014 and 2019, Cambodia experienced a notable decline in average trade costs with trading partners along the Belt and Road Initiative (BRI) corridors by 34.78 percent, twice as fast as with non-BRI trading partners. Regarding the decomposition of trade growth, we find that the expansion of Cambodian trade over the period from 1993 to 2019 was driven by three factors: the rise in income (59.65 percent), the decline in trade costs (56.69 percent), and the decline in multilateral resistance (–16.34 percent). The findings of this study have significant implications for a better understanding of Cambodia's development process toward global trade integration over the past two decades. Our results suggest that Cambodia can optimize its trade expansion potential by focusing on its relations with trading partners exhibiting high economic growth potential and those showing substantial reductions in trade costs.

## 1. Introduction

Trade costs have long been of great interest to trade economists owing to their critical roles in economic growth [1, 2], trade expansion [3, 4], trade networks [5], foreign direct investment (FDI) [6, 7], national welfare [8, 9], and economic development [9, 10]. In a broad sense, trade

stats.oecd.org/Index.aspx?DataSetCode=BTDIXE_I4#] and the United Nations Conference on Trade and Development (UNCTAD) database [https://unctadstat.unctad.org/datacentre/dataviewer/US.TradeMatrix]. All domestic trade data are available from the International Trade and Production Database for Estimation Release 2 (ITPD-E-R02) [https://www.usitc.gov/data/gravity/itpde.htm]. All Gross Domestic Product (GDP) data are accessible via the IMF's World Economic Outlook (WEO) database [https://www.imf.org/en/Publications/SPROLLs/world-economic-outlook-databases].

**Funding:** The author(s) received no specific funding for this work.

**Competing interests:** The authors have declared that no competing interests exist.

costs refer to all costs paid by consumers to obtain goods in addition to production costs [10]. According to the seminal Anderson and van Wincoop study [10], the tariff-equivalent trade costs of industrialized countries might be as high as 170 percent. This number can be further disaggregated into 55 percent for local distribution costs and 74 percent for international trade costs. Hummels et al. [11] argue that reducing a country's trade costs can increase vertical specialization and trade volumes, ultimately improving a country's position in global value chains (GVCs). Jacks et al. [4] found that a 16 percent decline in trade costs between 1950 and 2000 contributed almost 31 percent to the growth of global trade. These findings imply that despite increasing economic integration, trade costs continue to exert a profound impact on trade. In recent years, international organizations such as the World Trade Organization (WTO) and the Organization for Economic Co-operation and Development (OECD) have placed trade costs on their agendas. This inclusion is due to the significant global policy relevance of trade liberalization and trade facilitation for developing countries, especially least-developed countries (LDCs) [12].

Like in many other LDCs, high trade costs have long been a major obstacle to Cambodia's trade integration and liberalization. Inadequate infrastructure and inefficient transport and logistics facilities are the fundamental factors contributing to these high trade costs. Recent research indicates that Cambodia has the highest additional trade costs for inputs in specific industries, including textiles, chemicals, and computers, compared to other developing countries [13]. As of 2021, Cambodia's total imports comprised 59.87 percent of intermediate goods and 23.96 percent of consumer goods, compared to 19.87 percent of intermediate goods and 20.41 percent of consumer goods in the East Asia-Pacific region [14]. The fact that Cambodia heavily relies on imported intermediate and final goods implies that its high trade costs are likely to diminish the economic well-being of producers, suppliers, traders, and consumers.

On the policy side, the issue of trade costs has also been acknowledged in Cambodia's national policy agendas. For example, as highlighted by the Cambodia Trade Integration Strategy (CTIS) (2019–2023) in 2019 and the recent Pentagonal Strategy Phase-I (2023–2028) in 2023, high trade costs have been recognized as prominent trade barriers to be addressed in order to enhance Cambodia's international competitiveness and promote the diversity of its exports [15, 16]. The significance of export diversification and international competitiveness to support Cambodia's ambition to become an upper-middle-income country by 2030 and a high-income country by 2050 has been emphasized in many of its national policies and strategies. These include the Industrial Development Policy (IDP) (2015–2025) in 2015, the Rectangular Strategy-Phase IV (2018–2023) in 2018, the Cambodia Trade Integration Strategy (CTIS) (2019–2023) in 2019, the Economic Diplomacy Strategy (2021–2023) in 2021, and the Pentagonal Strategy-Phase I (2023–2028) in 2023, among various others. The literature suggests that high trade costs pose major challenges for a country to leverage local comparative advantages for export competitiveness [17], facilitate export diversification [18, 19], and engage in vertical specialization [11, 20]. For instance, Mora and Olabisi [19] found that high trade costs hinder developing countries' capacity to expand their new product lines or boost the export value of their current trade links. It is clear that reducing trade costs in Cambodia can play an indispensable role in its trade growth, economic integration, vertical specialization, export diversification, and international competitiveness.

Despite the extensive evidence highlighting the profound significance of trade costs in Cambodia from a practical and policy standpoint, little is known about their magnitudes, evolutions, and consequences. So far, the existing research on trade costs has made substantial contributions to deepening our understanding of their scales, determinants, and impacts [4, 10, 21]. However, to our knowledge, there have been hardly any empirical studies that exclusively examine trade costs in Cambodia. One of the main reasons for this is the difficulty in collecting reliable data on measuring trade costs, either directly or indirectly. While few studies have explored the factors affecting bilateral trade flows between Cambodia and its major trading partners using the traditional gravity model [22, 23], there has been a lack of an in-depth analysis of trade costs. For instance, using the traditional gravity model, Huot and Kakinaka [23] investigated the factors influencing Cambodia's bilateral trade flows with its top 20 trading partners from 2000 to 2004. Their study found that Cambodia's trade volumes decreased between 0.9 and 1.4 percent for every percentage point of increasing distance from its trading partners. The authors employed distance as a proxy for transport costs, which only captured a subset of trade costs. Therefore, the overall trade costs that have impeded Cambodia's trade flows might not be accurately measured. According to Anderson and van Wincoop [10], the overall trade costs cover all costs paid by consumers to obtain goods in addition to production costs. They include various elements, such as tariffs and non-tariff barriers, transport costs, border-related costs, language and communication barriers, and different currency effects, among others. In addition, as emphasized by Novy [24], conventional measures, such as distance and other static proxies for trade costs, cannot reflect the evolutions of trade barriers over time. These limitations in the literature highlight the need for an alternative approach that can determine Cambodia's comprehensive trade costs.

In light of the above discussions, this study aims to measure trade costs in Cambodia and explore the driving factors of its trade growth. To elaborate further, this research seeks to answer the following questions: What has been the trend in trade costs between Cambodia and its major trading partners over the past two decades? Which countries or regions have witnessed the most rapid decrease in their bilateral trade costs with Cambodia? What are the magnitudes of the remaining impediments? Are there any differences between Cambodia's average trade costs with developing countries and those with developed countries? Has the Belt and Road Initiative (BRI) reduced trade costs between Cambodia and its major trading partners? What factors have contributed to the expansion of Cambodia's trade? Is it due to changes in income or trade costs? To answer these questions, we employ a micro-founded method of trade costs introduced by Novy [24] to measure Cambodia's trade costs with its top 30 trading partners from 1993 to 2019. Unlike prior research in Cambodia that has solely relied on geographic distance as a proxy variable for transportation costs, this method allows us to track all costs that hinder the Cambodian trade flows without making assumptions about the individual components of these costs. We also apply the decomposition model of trade growth to examine the driving determinants of the Cambodian trade expansion during the same period.

This study contributes to the existing literature on trade costs in several ways. First, a deeper understanding of trade costs from LDCs, particularly low-income or lower-middle-income countries, has been limited [25–27]. Prior studies in this area have mainly focused on developed countries [24, 28], newly industrialized countries [29, 30], and a group of major developed and emerging economies [31, 32]. As highlighted by Anderson and van Wincoop [10], trade costs are much higher in developing countries and vary across countries. Therefore, this study offers new insights into how a least-developed country can mitigate trade barriers. In addition, earlier research findings on trade costs in developing countries have been inconclusive. For instance, some studies have shown a decline in trade costs over time [30, 33], whereas others have revealed no changes [29, 34, 35]. These conflicting results in the literature underscore the importance of conducting empirical studies to gain a better understanding of trade costs in developing countries that have also embraced trade liberalization. Next, the unique contribution of this research is that it is the first attempt to measure Cambodia's trade costs. Finally, this study presents novel data covering the 27-year period from 1993 to 2019, including Cambodia's top 30 trading partners, which accounted for nearly 94 percent of its overall trade volumes as of 2019. This extensive dataset allows us to track the dynamic changes in Cambodia's trade costs over time, from the initial stages of its economic reform in the early 1990s until the recent period, reflecting the country's trade openness with the world. By analyzing this data, we can also investigate the driving forces behind the growth of the Cambodian trade over the past two decades. Therefore, the findings will provide policymakers with the necessary information to target their policies to reduce trade costs and promote trade liberalization.

Our study presents multiple pieces of new evidence. First, Cambodia's average trade costs decreased by 35.43 percent between 1993 and 2019, with a remarkable decline of 24.66 percent between 2014 and 2019 alone. There were variations in these costs until 2014, even though the country joined the WTO in 2004. Next, Cambodia has lower bilateral trade costs with its major trading partners in Southeast Asia and East Asia than with those in South Asia, Oceania, Europe, and North America. Cambodia's average trade costs with developing and emerging economies are lower than those with developed economies. In terms of the magnitudes of these changes, Cambodia's average trade costs with developing and emerging economies have declined at twice the rate of those with developed economies since 2014. Notably, between 2014 and 2019, Cambodia's average trade costs with trading partners along the Belt and Road Initiative (BRI) corridors fell by 34.78 percent, double the rate of those with non-BRI trading partners. This finding aligns with the literature indicating that countries or regions along the BRI corridors where transport infrastructure projects have been developed experience a considerable decline in trade costs by up to 10.2 percent [36]. We also find that the expansion of Cambodian trade between 1993 and 2019 was driven by the rise in income (59.65 percent), the decline in trade costs (56.69 percent), and the decline in multilateral resistance (–16.34 percent). This study highlights the significance of reducing trade costs in supporting Cambodia's efforts to diversify its exports and enhance its international competitiveness. This will help Cambodia achieve its ambition to transform from a least-developed country to an upper-middle-income country by 2030 and a high-income country by 2050.

The remainder of this paper is outlined as follows. Section 2 provides an overview of Cambodia's trade development. Section 3 reviews the literature, presenting the theoretical background and empirical studies on trade costs. Section 4 contains the methodology for measuring trade costs and decomposing trade growth factors. Section 5 provides the empirical findings and discussions for Cambodia. Section 6 concludes. Finally, the S1 Appendix describes our data and sources.

## 2. Overview of Cambodia's trade development

Since adopting a market-based economy in 1993, Cambodia has implemented a series of major reform policies and strategies to boost exports, which are essential to economic development, employment, and poverty alleviation. One of the early national government policies was the first five-year Socio-Economic Development Plan (SEDP) of 1996–2000. Its primary goals were to achieve "high growth, employment generation, and poverty reduction", which were accomplished by promoting rural and labor-intensive industries through an export-led growth strategy [37]. Cambodia launched its first Diagnostic Trade Integration Strategy (DTIS) in 2001 after becoming a member of the Association of Southeast Asian Nations (ASEAN) in 1999. The adoption and implementation of these trade strategies made Cambodia the first LDC to join the WTO in 2004 through the complete accession procedure.

Consequently, Cambodia has witnessed remarkable growth in its trade. For example, Cambodia's exports and imports of merchandise increased from US$ 284 million and US$ 471 million in 1993 to US$ 14.82 billion and US$ 20.28 billion in 2019, respectively (see Fig 1). From

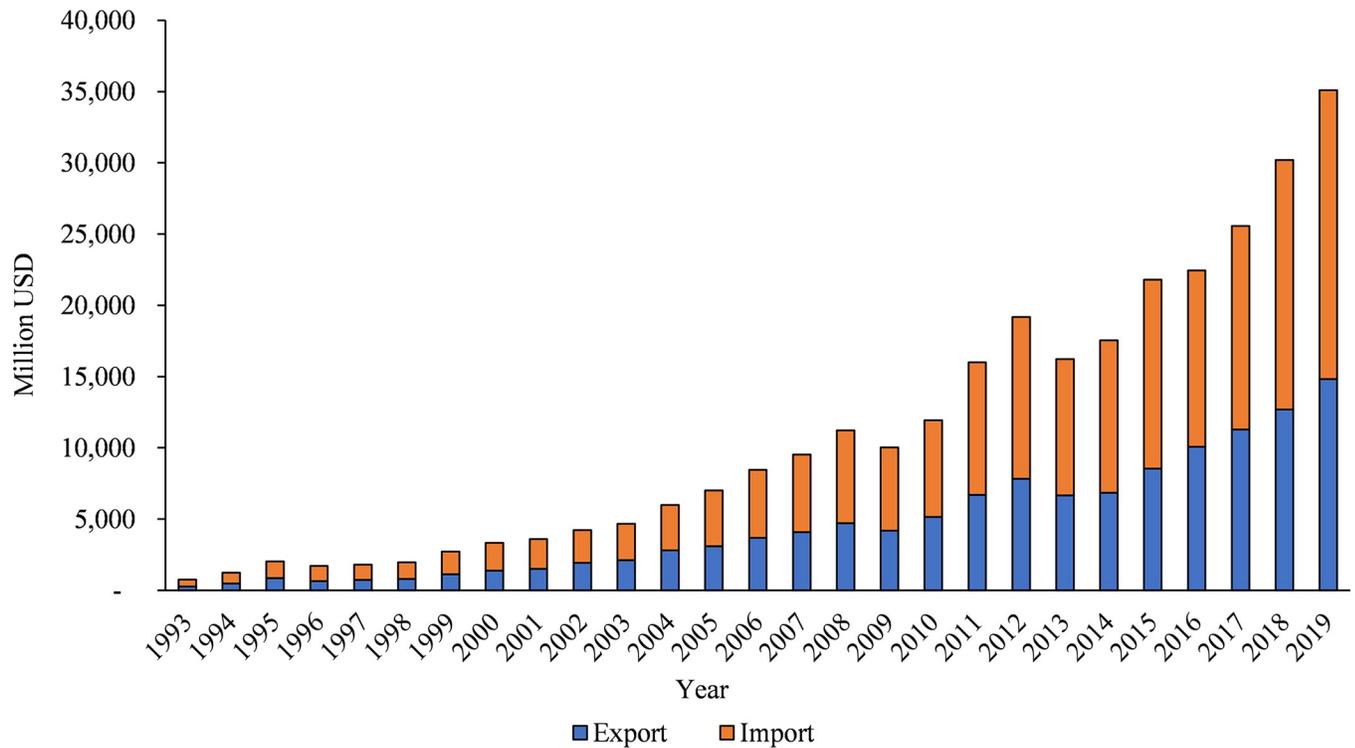

**Fig 1. Cambodia's merchandise trade volumes from 1993 to 2019 (in million USD). Source:** Authors' calculation using the merchandise export and import volumes, which are derived from the World Trade Organization (WTO) Stats data portal (https://stats.wto.org/).

https://doi.org/10.1371/journal.pone.0311754.g001

1993 to 2019, Cambodia recorded a 46.5-fold increase in its total merchandise trade volumes, rising from US$ 755 million to US$ 35.1 billion, equivalent to a growth rate of 15.9 percent per year. In comparison, the annual average growth rate of trade in goods in the ASEAN and the world during the same period stood at 7.3 percent and 6.3 percent, respectively. The average annual growth rate was computed by applying the compound annual growth rate from 1993 to 2019 for the total merchandise trade (export plus import) volumes in nominal terms, using the data from the WTO database [38]. It is evident from the data that Cambodia's trade witnessed an impressive growth rate over the period from 1993 to 2019, outperforming both the ASEAN and the global trade by more than double. Simultaneously, Cambodia's merchandise trade as a share of its gross domestic product (GDP) increased from 29.8 percent in 1993 to 129.6 percent in 2019 [39]. In 2016, Cambodia became the second-largest exporter of manufactured goods among the LDCs after Bangladesh, with total exports reaching more than US$ 10 billion [40]. This rapid growth in Cambodian trade over the past two decades can be attributed to various factors, including a decrease in both tariff and non-tariff barriers, the signing of more regional trade agreements (RTAs), improvements in information and transport technology, better infrastructure development, a growth in foreign direct investment (FDI), and an increase in income. Previous studies have verified that reducing trade costs is necessary for expanding international trade [3, 4, 24]. For instance, Jacks et al. [4] found that the growth of global trade between 1950 and 2000 was contributed by nearly 31 percent due to a decline in trade costs.

Cambodia had been recognized as one of the most rapidly growing economies in the world until the COVID-19 pandemic hit in 2020. The Cambodian economy grew by 7.7 percent annually on average between 1995 and 2019, while the per capita income increased from US$

323 in 1995 to US$ 1,621 in 2019, allowing Cambodia to attain its lower-middle-income status in 2015 [41]. This remarkable economic development in Cambodia over the past 25 years would not have been possible without the rapid growth of its trade. Despite all these achievements, Cambodia's exports have been concentrated largely on low-value-added products, including textiles, clothing, footwear, travel accessories, and rice. Over the past decade, more than half of Cambodia's exports in goods have gone to the United States and the European Union (EU) markets [42]. This heavy dependence on a few developed countries makes Cambodia's exports vulnerable to changes in their market demand. As an illustration, Cambodia recorded a decline in exports in 2009 because of the 2008–2009 Global Financial Crisis and again in 2013 and 2014, partly due to the low import demand in developed countries and the recession in the European Union (see Fig 1) [43]. The other concern that needs to be considered is that if Cambodia graduates from its LDC status, it will no longer be eligible for trade preferences or measures related to international assistance under the Generalized System of Preferences (GSP).

The empirical investigation has demonstrated that Cambodian exports would decrease by up to 10 percent upon transitioning from its LDC status [44]. It is clear that the conventional approach of depending on exports of low-value-added products to boost Cambodia's economic growth is no longer sustainable. Therefore, a new development paradigm is necessary to diversify Cambodia's export markets or upgrade its current low-value-added products to medium- or high-value-added products. In this sense, the Royal Government of Cambodia (RGC) is fully aware of these challenges and has actively integrated the country's trade and economic activities through various bilateral, regional, and global measures, such as trade facilitation initiatives, economic partnership agreements, and regional trade agreements (RTAs). These include the Belt and Road Initiative (BRI) proposed by China in 2013, the Cambodia-Hong Kong (China) Free Trade Agreement in February 2021 under the ASEAN-Hong Kong (China) Free Trade Agreement (AHKFTA), the Regional Comprehensive Economic Partnership (RCEP) in January 2022, the Cambodia-China Free Trade Agreement (CCFTA) in January 2022, the Cambodia-Republic of Korea Free Trade Agreement (CKFTA) in December 2022, and the Cambodia-United Arab Emirates Comprehensive Economic Partnership Agreement (CAM-UAE CEPA) in 2023. It is important to note that the RCEP, officially coming into effect on January 1, 2022, is the world's largest free trade agreement (FTA) in terms of its economic and population size [45]. This trade block comprises 15 countries in the Asia-Pacific regions, including five countries of the ASEAN Plus Six (Australia, China, Japan, New Zealand, and the Republic of Korea) and ten countries of the ASEAN (Brunei, Cambodia, Indonesia, Laos, Malaysia, Myanmar, the Philippines, Singapore, Thailand, and Vietnam). In recent years, Cambodia has actively tried to negotiate and propose several economic partnership agreements and RTAs with Canada, India, Japan, Switzerland, and the Eurasian Economic Union (EAEU). These efforts aim to reduce Cambodia's trade costs with its major trading partners and diversify its export markets.

## 3. Literature review

### 3.1 Theoretical review of trade costs

This section presents an overview of the theoretical background of trade costs and their measurement methods. Since Paul A. Samuelson [46] introduced the concept of modeling transport costs as "iceberg transport costs" in 1954, trade costs have received significant attention from trade economists in determining trade volumes. However, the role of trade costs in shaping trade patterns has not always been a prominent focus in traditional trade theories. As Deardorff [47] emphasized, many of the earlier trade theories assumed the absence of trade costs, which led to an incomplete understanding of the factors influencing the trade volumes of a country. As international trade theories have evolved, trade costs have emerged as crucial components in prominent theories, such as the new trade theory [48, 49], the new economic geography theory [50, 51], and the theory of heterogeneous firms [52, 53]. Trade costs are also critical in resolving major empirical puzzles within international macroeconomics, such as the home bias in trade, the consumption correlations puzzle, and the purchasing power parity (PPP) puzzle [54]. We adopt a comprehensive definition of trade costs proposed by Anderson and van Wincoop [10]: "Trade costs, broadly defined, include all costs incurred in getting a good to a final user other than the marginal cost of producing the good itself: transportation costs (both freight costs and time costs), policy barriers (tariffs and nontariff barriers), information costs, contract enforcement costs, costs associated with the use of different currencies, legal and regulatory costs, and local distribution costs (wholesale and retail)". According to Jacks et al. [3], trade costs are related to transactions and transportation costs that hinder international economic integration in exchanging goods across borders.

There are two approaches to measuring trade costs: direct measurement [55, 56] and indirect measurement [10, 57, 58]. The direct measurement method mainly selects specific indicators to represent trade costs, such as transportation costs, information costs, tariffs, and nontariff barriers (see Chen and Novy [59] for a brief discussion on the measurement method of trade costs). Although direct measures are straightforward, they are constrained by the unavailability of data, especially for developing countries. They can only be used to quantify specific subsets of trade costs instead of the overall trade costs. Measuring the overall trade costs is challenging since "direct measures are remarkably sparse and inaccurate" [10]. Given the limitations of direct approaches in measuring other components of trade costs, scholars have opted for indirect methods through price data and trade volumes (see [10] for a full review of the trade cost measure using price data). The standard gravity model, the workhorse trade model pioneered by Jan Tinbergen in 1962 [60], has been traditionally used to infer trade costs indirectly from trade flows, with particular variables like geographic distance as a proxy for transportation costs. Other commonly used proxy variables to measure different types of trade costs are currency barriers [61], transportation and infrastructure barriers [55, 56, 62], tariff barriers [62], and language and communication barriers [55, 63, 64], among others. Other scholars have used border-related trade barriers to quantify trade costs [57, 58, 64–66]. Overall, there are two issues with all these measuring techniques. First, they might not entirely reflect all barriers that hinder trade because they cannot capture other hidden trade cost variables, such as cultural differences, institutional barriers, and bureaucratic red tape, which can be difficult to measure using proxy variables. Second, researchers typically determine variables related to trade costs beforehand and incorporate them into the gravity model for regression analyses, which can lead to biased results due to omitted variables. To fill these theoretical gaps, Novy [24], based on Head and Ries [66], developed a micro-founded equation of trade costs that can capture all costs associated with exchanges of goods between two countries or regions without making assumptions about the components of these costs.

### 3.2 Empirical studies on trade costs

In this section, we present a review of empirical studies on trade costs to highlight the importance of our contribution to the existing literature. In terms of contributions to empirical studies on trade costs, Anderson and van Wincoop [10] estimated trade costs in developed countries, and their findings revealed that the tariff equivalent of trade costs amounted to approximately 170 percent. This number can be further disaggregated to 21% transportation costs, 44% border-related trade costs, and 55% sales (retail and wholesale distribution) costs.

Novy [24] measured trade costs between the United States and its major trading partners from 1970 to 2000, and the study found that the average tariff equivalent trade costs of the United States declined by 40 percent throughout the sample period. According to the same survey, the average trade costs of the 13 OECD countries decreased by about one-third between 1970 and 2000. Subsequently, the literature has extensively adopted Novy's trade cost equation to estimate comprehensive trade costs between countries or regions. For example, Wen et al. [34] used Novy's trade cost model to measure the bilateral trade costs of China's agricultural goods with its five major trading partners from 1995 to 2007. Their findings revealed that China's trade costs in agricultural goods had not decreased. Similarly, Miroudot et al. [33] employed an identical approach to estimate trade costs in goods and services, and the authors found that trade costs were much higher in services than goods. Their study also indicated a decrease of approximately 15 percent in global aggregate trade costs in goods between 1995 and 2007.

Arvis et al. [27] adopted the same metric to analyze trade costs in manufacturing and agricultural goods in developing countries from 1996 to 2010. The authors concluded that agricultural trade costs were twice as high as those in manufactured goods and were extremely high for low-income countries. According to the same study, only upper-middle-income countries could reduce trade costs faster among developing countries, while trade costs in low-income countries were still very high. Gaurav and Mathur [29] analyzed the bilateral trade costs between India and the European Union trading partners from 1995 to 2010. Their study revealed that average trade costs between India and the European Union steadily decreased until 2001. After that, these trade costs remained stable at around 50 percent for nearly a decade. Turkson [35] also employed Novy's equation to estimate trade costs in 39 sub-Saharan African countries from 1980 to 2003. The author discovered that Sub-Saharan Africa had the highest average trade costs at 271.5 percent, compared with other regions. In addition, Sub-Saharan Africa had not experienced a decline in trade costs during the same period.

In summary, the existing literature on trade costs has the following characteristics. First, previous studies in this area have primarily focused on developed countries, newly industrialized countries, and groups of developed and emerging countries. However, there has been limited research on trade costs in developing countries, particularly the LDCs. Second, early research findings from developing countries have shown significant contradictions. For instance, some studies have discovered a decrease in trade costs over time, whereas others have revealed no changes. Overall, the existing literature has concluded that trade costs tend to decrease among major developed nations. However, this trend does not apply to developing countries, especially the LDCs, where trade costs remain very high or show no signs of decline.

Being the first LDC to join the WTO in 2004, Cambodia's endeavor to reduce its trade barriers has yet to be fully understood, primarily from the perspective of trade costs. Reducing trade costs is even more critical and urgent for LDCs like Cambodia, seeking to gain comparative advantages and increase export competitiveness regionally and globally. Deardorff [17] highlighted that the local comparative advantages determine the competitiveness of export commodities. However, when trade costs increase, the local comparative advantages decrease, thereby restricting the country's ability to enhance its export competitiveness. The importance of trade costs emphasizes the need for an in-depth understanding of their scales, evolutions, and impacts. Unfortunately, there have been hardly any empirical studies that exclusively examine trade costs in Cambodia. Earlier empirical studies on Cambodian trade were conducted by scholars like Kim [22] and Huot and Kakinaka [23], who were among the first to use the traditional gravity model to measure bilateral trade flows between Cambodia and its major trading partners. Despite the valuable contributions of their studies in understanding Cambodia's trade openness before and after its membership in the ASEAN (1999) and the WTO (2004), the investigation of Cambodia's trade costs, including their magnitudes, evolutions, and consequences, has remained largely unexplored. For instance, using the traditional gravity model, Huot and Kakinaka [23] examined the factors influencing Cambodia's bilateral trade flows with its top 20 trading partners from 2000 to 2004. Their study found that Cambodia's trade volumes decreased between 0.9 and 1.4 percent for every percentage point of increasing distance from its trading partners. The authors employed distance as a proxy variable for transportation costs, which might only reflect a portion of the overall trade costs. Therefore, the comprehensive trade costs that hinder Cambodia's trade flows, such as border-related costs, logistics costs, and costs associated with trade compliance, can not be fully captured. Moreover, the gravity model can only explain how distance affects the Cambodian trade volumes. However, it does not provide details about the scales of transportation costs or how these costs have evolved. As Novy [24] pointed out, time-invariant proxy variables, such as distance, are not well-suited to reflect the dynamic variations of trade costs over time. Finally, the traditional gravity model does not take into account the impact of multilateral resistance on trade, leading to biased results [58].

Against this background, this study aims to fill these gaps in the existing literature on trade costs in the context of LDCs, particularly in the case of Cambodia. Using Novy's micro-founded equation of trade costs, this study measures Cambodia's trade costs with its top 30 trading partners from 1993 to 2019. We also employ the trade growth decomposition model to investigate the driving forces behind Cambodia's trade growth during the same period. This study provides fresh insights into how Cambodia's trade costs have evolved over the past two decades after the country transformed into a market-based economy in the early 1990s.

## 4. Material and method

### 4.1 Micro-founded measure of trade costs

**4.1.1 Measurement method of bilateral trade costs.** Building upon the famous border trade puzzle between the United States and Canada presented by McCallum [57], Anderson and van Wincoop [58] made a seminal contribution to the gravity model by including a multilateral resistance component and addressing potential bias resulting from omitted variables. The authors propose that the bilateral trade cost equation between countries or regions is a mathematical function of border costs and geographic distance between trading partners. They define the bilateral trade costs ($t_{ij}$) as a log-linear function $t_{ij} = b_{ij}\, d_{ij}^{\rho}$, where $b_{ij}$ denotes the border-related variables ($b_{ij} = 1$ if both regions $i$ and $j$ are in the same country; otherwise, $b_{ij}$ is the sum of one and the tariff equivalent of the borders separating the regions' respective countries); $d_{ij}$ represents distance between the two countries or regions, and $\rho$ is the elasticity of distance. First, since trade barriers vary across countries, Novy [24] raises concerns about the oversimplification of Anderson and van Wincoop's model by assuming that trade costs between countries are symmetric ($t_{ij} = t_{ji}$). Next, the trade cost function might not be accurately defined because it might exclude significant trade cost factors like tariffs. Finally, he emphasizes that static measures, such as geographic distance, cannot capture the variations in trade costs over time. To overcome these shortcomings, Novy fills theoretical and analytical gaps by deriving a micro-founded equation of trade costs based on Anderson and van Wincoop's gravity model. Novy's proposed measure of bilateral trade costs has several major advantages. First, it can calculate all the costs involved in trading goods between two countries or regions without making assumptions about the components of these costs. Second, this method applies not only to cross-sectional data but also to time series and panel data. Third, the micro-founded measure of trade costs exhibits robust theoretical foundations and can be derived from other influential trade models, such as the Ricardian model of comparative advantage [64] and the heterogeneous firms' models [67, 68]. Inspired by these substantial improvements over the previous models, we follow Novy's equation for measuring the trade costs of Cambodia.

The following gravity equation, derived by Anderson and van Wincoop [58], has the form illustrated in Eq (1):

$$X_{ij} = \left(\frac{Y_i Y_j}{Y^W}\right)\left(\frac{t_{ij}}{\Pi_i P_j}\right)^{1-\sigma} \quad (1)$$

Where $X_{ij}$ denotes the nominal export volume from country $i$ to country $j$; $Y_i$ and $Y_j$ represent the nominal income of countries $i$ and $j$; $Y^W$ is the world's nominal income, defined as $Y^W \equiv \Sigma_j Y_j$, $t_{ij} \geq 1$ denotes the bilateral trade costs (an ad valorem term plus one), and $\sigma > 1$ is the elasticity of substitution between commodities. $\Pi_i$ and $P_j$ represent the price indices of country $i$ and country $j$, respectively. Anderson and van Wincoop call the two terms "multilateral resistance" to describe these price indices, which can be defined as average trade costs with all trading partners. $\Pi_i$ is the outward multilateral resistance of country $i$, while $P_j$ is the inward multilateral resistance of country $j$.

Considering the symmetry of countries $i$ and $j$ in Eq (1), the gravity equation of the bilateral export from country $j$ to country $i$, denoted as $X_{ji}$, can be written:

$$X_{ji} = \left(\frac{Y_j Y_i}{Y^W}\right)\left(\frac{t_{ji}}{\Pi_j P_i}\right)^{1-\sigma} \quad (2)$$

This study refers to the method proposed by Head and Ries [66] and Novy [24] to eliminate multilateral resistance variables.

Following Novy's proposal, we extend Eq (1) to express the domestic trade volumes of country $i$ ($X_{ii}$):

$$X_{ii} = \left(\frac{Y_i Y_i}{Y^W}\right)\left(\frac{t_{ii}}{\Pi_i P_i}\right)^{1-\sigma} \quad (3)$$

In Eq (3), $\tau_{ii}$ represents the domestic trade costs of country $i$. By rewriting Eq (3), we can obtain the product of the outward multilateral resistance and the inward multilateral resistance of country $i$:

$$\Pi_i P_i = t_{ii} \left(\frac{X_{ii}/Y_i}{Y_i/Y^W}\right)^{\frac{1}{(\sigma-1)}} \quad (4)$$

Similarly, the product of the outward multilateral resistance and the inward multilateral resistance of country $j$ can be obtained:

$$\Pi_j P_j = t_{jj} \left(\frac{X_{jj}/Y_j}{Y_j/Y^W}\right)^{\frac{1}{(\sigma-1)}} \quad (5)$$

We multiply the two sides of Eq (1) and Eq (2) to arrive at a two-way gravity equation:

$$X_{ij} X_{ji} = \left(\frac{Y_i Y_j}{Y^W}\right)^2 \left(\frac{t_{ij} t_{ji}}{\Pi_i P_i \Pi_j P_j}\right)^{1-\sigma} \quad (6)$$

We substitute Eqs (4) and (5) into Eq (6) to get:

$$\left(\frac{t_{ij}t_{ji}}{t_{ii}t_{jj}}\right) = \left(\frac{X_{ii}X_{jj}}{X_{ij}X_{ji}}\right)^{\frac{1}{(\sigma-1)}} \tag{7}$$

Considering the asymmetry of trade costs between country *i* and country *j* ($t_{ij} \neq t_{ji}$), their domestic trade costs are also different ($t_{ii} \neq t_{jj}$). Following Novy's approach, we get the equation of trade costs in tariff equivalent by calculating the geometric mean of trade costs in all directions and then subtracting one from both sides in Eq (7). It should be noted that the derivation of a micro-founded measure of trade costs has been done by Novy, Dennis (see Novy [24] for the details of the discussion). This approach is similar to the models proposed by Head and Ries [66] and Head and Mayer [69].

$$\tau_{ij} = \left(\frac{t_{ij}t_{ji}}{t_{ii}t_{jj}}\right)^{\frac{1}{2}} - 1 = \left(\frac{X_{ii}X_{jj}}{X_{ij}X_{ji}}\right)^{\frac{1}{2(\sigma-1)}} - 1 \tag{8}$$

In Eq (8), $\tau_{ij}$ denotes the tariff equivalent trade costs, which measure the bilateral trade costs of the two trading partners ($t_{ij}t_{ji}$) relative to their respective domestic trade costs ($t_{ii}t_{jj}$). Assuming all other variables remain constant, Eq (8) illustrates that if the bilateral trade ($X_{ij}X_{ji}$) of the two trading partners grows faster than their respective domestic trade ($X_{ii}X_{jj}$), trade costs will decrease.

**4.1.2 Measurement method of average trade costs.** This portion refers to the method derived by Arvis et al. [27] to compute the annual average trade costs between Cambodia and all of its trading partners. The geometric average of the actual total trade between country *i* and its trading partner *j* can be defined as

$$\bar{X}_{ij} = \bar{X}_{ji} = (X_{ij} \times X_{ji})^{\frac{1}{2}}. \tag{9}$$

The authors set the following equation, based on the bilateral trade cost model of Novy (2013), to get the formula for calculating average trade costs ($\bar{\tau}_t$):

$$\bar{\tau}_t = \left(\frac{\sum_{j \neq i} \bar{X}_{ij}}{(X_{ii})^{\frac{1}{2}} \sum_{j \neq i} (X_{jj})^{\frac{1}{2}}}\right)^{\frac{1}{1-\sigma}} - 1 \tag{10}$$

In Eq (10), the annual average trade costs can be measured using the domestic trade data of both countries, the elasticity of substitution between goods, and the geometric average of the actual total trade as defined in Eq (9). $\bar{\tau}_t$ represents the annual average trade costs of Cambodia with all of its trading partners in the given year *t* (*t* = 1993, 1994,..., 2019).

## 4.2 Decomposition model of trade growth

This sub-section presents an overview of the existing literature concerning the factors that contribute to the expansion of trade and introduces the decomposition model of trade growth. Krugman [70] and Feenstra [71] argue that one of the significant drivers of trade expansion is the decline in trade costs. Baier and Bergstrand [62] investigated the factors influencing bilateral trade in 16 OECD countries between 1958 and 1988. Their findings indicated that income, tariffs, and transportation costs had significant impacts on trade growth, with income growth having the largest impact (67%), followed by a reduction in tariffs (25%) and a decrease in transportation costs (8%). Jacks et al. [21] decomposed the growth of global trade based on two main determinants: income and trade costs. They found that the growth of global trade between 1870 and 1913 was driven by 44 percent due to a reduction in trade costs. Jacks et al. [4] discovered that the decline in trade costs accounted for almost 31 percent of the global trade growth between 1950 and 2000. In summary, previous studies have confirmed that reducing trade costs is necessary for expanding international trade.

So far, there has been a lack of research on the determinants of trade growth in Cambodia. Hence, the question arises: has the decline in trade costs played a role in Cambodia's trade growth over the past two decades? In order to answer this question, this section decomposes the trade growth equation derived by Novy [24] into three components: income, trade costs, and multilateral resistance.

First, we take the natural logarithm for both sides of Eq (6). Then, we derive the first-order difference of the natural logarithm result to obtain Eq (11):

$$\Delta\ln(X_{ij}X_{ji}) = 2\Delta\ln\left(\frac{Y_iY_j}{Y^W}\right) + (1-\sigma)\Delta\ln(t_{ij}t_{ji}) - (1-\sigma)\Delta\ln(\Pi_iP_i\Pi_jP_j) \quad (11)$$

Eq (11) indicates that the bilateral trade growth, $\Delta\ln(X_{ij}X_{ji})$, is influenced by three factors: the first is the change in each country's income compared to the world's income, $\Delta\ln\left(\frac{Y_iY_j}{Y^W}\right)$; the second is the change in the bilateral trade costs, $\Delta\ln(t_{ij}t_{ji})$; and the third is the change in the level of multilateral resistance between trading partners, $\ln(\Pi_iP_i\Pi_jP_j)$. Since the value of the bilateral trade costs $t_{ij}$ and $t_{ji}$ are unknown, we substitute Eq (8) into Eq (11) to obtain the following Eq (12):

$$\Delta\ln(X_{ij}X_{ji}) = 2\Delta\ln\left(\frac{Y_iY_j}{Y^W}\right) + 2(1-\sigma)\Delta\ln(1+\tau_{ij}) - 2(1-\sigma)\Delta\ln(\Phi_i\Phi_j) \quad (12)$$

Among them, $\Phi_i = \left(\frac{\Pi_iP_i}{t_{ii}}\right)^{1/2}$ and $\Phi_j = \left(\frac{\Pi_jP_j}{t_{jj}}\right)^{1/2}$

In Eq (12), $\Phi_i$ and $\Phi_j$ denote the multilateral resistance of country $i$ and country $j$, respectively, relative to their domestic trade costs.

Further, we divide both sides of Eq (12) by $\Delta\ln(X_{ij}X_{ji})$ to obtain the bilateral trade decomposition equation as follows:

Fig 2 explains that the growth of bilateral trade between the two trading partners comes from the contribution of three determinants: (A) the growth of income, (B) the reduction in bilateral trade costs, and (C) the decrease in multilateral resistance. Assuming bilateral trade barriers were constant over time between both countries, the combined contribution of (B) and (C) would amount to zero. As a result, the growth of income (A) would be the sole contributor to trade growth. If bilateral trade costs decrease (i.e., $\Delta\ln(1+\tau_{ij})<0$), it indicates that the contribution of the decline in bilateral trade costs to the growth of bilateral trade is positive, inferring that (B) is positive. If the multilateral resistance declines (i.e., $\Delta\ln(\Phi_i\Phi_j)<0$), then (C) becomes negative. This negative impact can be explained as the trade diversion effect due to a

$$100\% = \underbrace{\frac{2\Delta\ln\left(\frac{Y_iY_j}{Y^W}\right)}{\Delta\ln(X_{ij}X_{ji})}}_{(A)} + \underbrace{\frac{2(1-\sigma)\Delta\ln(1+\tau_{ij})}{\Delta\ln(X_{ij}X_{ji})}}_{(B)} - \underbrace{\frac{2(1-\sigma)\Delta\ln(\Phi_i\Phi_j)}{\Delta\ln(X_{ij}X_{ji})}}_{(C)} \quad (13)$$

**Fig 2. Eq 13.**

https://doi.org/10.1371/journal.pone.0311754.g002

decline in multilateral resistance. It can also be interpreted that when barriers to international trade decrease with other countries, there is a subsequent increase in trade between those countries, resulting in a decline in bilateral trade between countries $i$ and $j$.

To solve $2(1-\sigma)\Delta\ln(1+\tau_{ij})$, we recall Eq (8).

$$2(1-\sigma)\Delta\ln(1+\tau_{ij}) = \Delta\ln(X_{ij}X_{ji}) - \Delta\ln(X_{ii}X_{jj}) \tag{14}$$

To solve $2(1-\sigma)\Delta\ln(\Phi_i\Phi_j)$, we recall Eqs (4) and (5).

$$2(1-\sigma)\Delta\ln(\Phi_i\Phi_j) = \Delta\ln\left(\frac{Y_i/Y^W}{X_{ii}/Y_i}\right) + \Delta\ln\left(\frac{Y_j/Y^W}{X_{jj}/Y_j}\right) \tag{15}$$

The decomposition equation exemplifies that the growth of bilateral trade does not depend on the value of the elasticity of substitution between goods ($\sigma$).

By substituting Eqs (14) and (15) into Fig 2, we can obtain the following Eq (16):

$$100\% = \frac{2\Delta\ln\left(\frac{Y_iY_j}{Y^W}\right)}{\Delta\ln(X_{ij}X_{ji})} + \frac{\Delta\ln(X_{ij}X_{ji}) - \Delta\ln(X_{ii}X_{jj})}{\Delta\ln(X_{ij}X_{ji})} - \frac{\Delta\ln\left(\frac{Y_i/Y^W}{X_{ii}/Y_i}\right) + \Delta\ln\left(\frac{Y_j/Y^W}{X_{jj}/Y_j}\right)}{\Delta\ln(X_{ij}X_{ji})} \tag{16}$$

### 4.3 Sample, data sources, and parameter settings

**4.3.1 Sample.** In this study, the top 30 trading partners are selected based on the availability of data and their stable bilateral trade volumes with Cambodia over the past two decades. In 2019, all these trading partners collectively made up approximately 94 percent of Cambodia's total exports and imports, based on the data from the International Monetary Fund (IMF)'s Direction of Trade Statistics (DOTS) [72]. As of 2019, Cambodia and its top 30 trading partners contributed to over 76 percent of the world's total exports and imports, 73 percent of the world's economy measured by the purchasing power parity (PPP), and more than half of the global population [72, 73]. Therefore, this representative sample can be used to infer Cambodia's overall trade pattern with the world. These top 30 trading partners include 13 countries in the G20 (Australia, Canada, China, Germany, France, India, Indonesia, Italy, Japan, the Republic of Korea, the Russian Federation, the United Kingdom, and the United States), eight countries in the European Union (EU) that are not the members of the G20 (Austria, Belgium, Denmark, Finland, Ireland, the Netherlands, Spain, and Sweden), five countries in the RCEP, that are not the members of the G20 (Malaysia, New Zealand, Singapore, Thailand, and Vietnam), and other major trading partners (Hong Kong (China), Norway, Switzerland, and the Taiwan Province of China). This paper uses the annual data for the 27-year period spanning from 1993 to 2019, equivalent to a total of 810 observations ($N = 30 \times T = 27$). It is important to note that we select the post-1993 sample to highlight the development of Cambodia's trade liberalization, as the country has only started official trade with many countries and regions since implementing a market-oriented economy in 1993.

**4.3.2 Data sources.** The primary source of the bilateral export data in free-on-board (FOB) is derived from the IMF's Direction of Trade Statistics (DOTS) database. However, the data of some trading partners are not available from the IMF's DOTS database. Therefore, where possible, the missing data are supplemented using the data from the OECD's Structural Analysis (STAN) database and the United Nations Conference on Trade and Development (UNCTAD) database (refer to the S1 Appendix for the detailed descriptions of this procedure). The domestic trade data are based on the latest update (second edition) dataset from the International Trade and Production Database for Estimation Release 2 (ITPD-E-R02), published in

Table 1. Description of the variables and data sources.

| Variable | Description | Sources |
| --- | --- | --- |
| $X_{ij}$ | Nominal export from Cambodia to country $j$ (Thousand USD) | IMF's DOTS, OECD's STAN and UNCTAD |
| $X_{ji}$ | Nominal export from country $j$ to Cambodia (Thousand USD) | |
| $X_{ii}$ | Domestic trade of Cambodia (Thousand USD) | ITPD-E-R02 |
| $X_{jj}$ | Domestic trade of country $j$ (Thousand USD) | |
| $Y_i$ | Nominal income of Cambodia (Thousand USD) | IMF's WEO |
| $Y_j$ | Nominal income of country $j$ (Thousand USD) | |
| $Y^W$ | World's nominal income (Thousand USD) | |

**Source:** Authors' Elaboration

https://doi.org/10.1371/journal.pone.0311754.t001

July 2022. The ITPD-E-R02 database covers the data for 265 countries or regions and 170 industries in the four sectors, including agriculture, mining & energy, manufacturing, and services, from 1986 to 2019 [74, 75]. It is important to emphasize that the scope of this research is limited to trade in the goods sector and total trade costs. Due to the unavailability of data on domestic trade in other sectors for Cambodia and other developing countries, this research does not analyze disaggregated trade costs by industries. To obtain the total domestic trade data in the goods sector, we aggregate the three industries, including agriculture, mining & energy, and manufacturing, from the ITPD-E-R02 database. The GDP data are also needed to decompose the growth of the Cambodian trade using Eq (16), and they are taken from the IMF's World Economic Outlook (WEO) database published in October 2022. In order to capture the price indices in what Anderson and van Wincoop [58] refer to as "multilateral resistance", trade volumes and income are estimated using the nominal value. All the data are measured in thousands of current US dollars for the corresponding year. Table 1 summarizes the variables used to measure trade costs, their descriptions, and their data sources, and Table 2 presents the descriptive statistics of these variables.

**4.3.3 Parameter settings.** To calculate trade costs using Eqs (8) and (10), it is necessary to make the parameter assumption. Specifically, determining the exact value of the elasticity of substitution across goods is challenging. Based on a summary of estimated results from various existing literature, as reported by Anderson and van Wincoop [10], the elasticity of substitution across goods ($\sigma$) is likely to fall between 5 and 10. Novy [24] demonstrates that changes in trade costs are not affected by changes in the elasticity of substitution. He observes that between 1970 and 2000, when $\sigma = 10$, the average bilateral trade costs of the United States

Table 2. Descriptive statistics of the variables.

| Variable | Observations | Mean value | Standard deviation | Minimum value | Maximum value |
| --- | --- | --- | --- | --- | --- |
| $X_{ij}$ | 807 | 148244.6 | 371330 | 0.1113 | 4414284 |
| $X_{ji}$ | 807 | 277875.2 | 830814.8 | 7.153 | 8000644 |
| $X_{ii}$ | 810 | 3009412 | 2335690 | 618366.1 | 7859559 |
| $X_{jj}$ | 810 | 4.97E+08 | 1.42E+09 | 5257 | 1.61E+10 |
| $Y_i$ | 810 | 9968074 | 7276862 | 2427000 | 2.71E+07 |
| $Y_j$ | 810 | 1.48E+09 | 2.82E+09 | 1.67E+07 | 2.14E+10 |
| $Y^W$ | 810 | 5.41E+10 | 2.07E+10 | 2.61E+10 | 8.77E+10 |

**Source:** Authors' Calculation

https://doi.org/10.1371/journal.pone.0311754.t002

decreased from 54 percent to 31 percent, a decline of 42 percent; when $\sigma$ = 8, trade costs decreased from 74 percent to 42 percent, a decline of 44 percent. Therefore, in this paper, we set the value of the elasticity of substitution across products ($\sigma$) to 8, as suggested by the previous studies [10, 24, 27].

## 5. Empirical results and discussions

### 5.1 Measurement results and analyses of bilateral trade costs between Cambodia and major trading partners

In this section, we present the results of the ad valorem equivalent bilateral trade costs between Cambodia and 30 of its top trading partners between 1993 and 2019.

Table 3 presents the levels and the percentage changes in the bilateral trade costs (ad valorem equivalent) between Cambodia and 30 of its top trading partners between 1993 and 2019. The trading partners are organized in ascending order based on their relative bilateral trade costs with Cambodia as of 2019 across regions. There are significant variations in the absolute value of the bilateral trade costs across trading partners. For instance, Cambodia's bilateral trade costs with Hong Kong (China) were as low as 1.35% in 2019, while this number reached 251.25% with Norway in the same year. Notably, Norway is the only trading partner with trade costs exceeding 200%. Over the years from 1993 to 2019, trade costs of Hong Kong (China), Thailand, Vietnam, and Singapore remained below 100%, which were consistently lower than those of other trading partners. We offer a few possible explanations for this phenomenon. First, Hong Kong (China) and Singapore function as the major entrepot hubs in the Asia-Pacific region, facilitating substantial re-export trade. This means that their domestic trade flows are relatively smaller than their foreign trade flows, leading to lower trade costs. Second, Hong Kong (China) and Singapore boast some of the lowest tariffs and exhibit a high degree of trade openness. Third, both trading partners are geographically close to Cambodia, resulting in low transportation costs. Most importantly, Singapore has shared the Free Trade Agreement (FTA) with Cambodia under the ASEAN Free Trade Area (AFTA) since 1999, which is vital for reducing trade barriers between the two countries. Refer to the following section for a comprehensive discussion of the bilateral trade costs between Cambodia and its neighboring countries (Thailand and Vietnam).

In terms of the regional analysis, except for Indonesia in Southeast Asia, all trading partners whose trade costs were over 100% in 2019 are situated in South Asia, Oceania, Europe, and North America, with the majority being in Europe. In general, the trade costs of the short-distance trading partners in Southeast Asia and East Asia are lower than most of the long-distance counterparts in Europe and North America. This finding infers that geographic distance is one of the most critical factors affecting trade costs, which aligns with the theoretical framework of the gravity model. The hypothesis of this model suggests that trading partners located far from each other tend to engage in less trade due to high transportation costs [58, 60].

Another notable finding is that despite being a part of the ASEAN Free Trade Area (AFTA) with Indonesia since 1999, the ASEAN-Australia-New Zealand Free Trade Area (AANZFTA) with Australia and New Zealand since 2010, and the ASEAN-India Free Trade Area (AIFTA) with India since 2010, Cambodia's bilateral trade costs with these countries have remained relatively high during the specific time period. Therefore, it is essential to focus on ongoing efforts to reduce trade barriers and improve trade facilitation, particularly in the context of the recent Regional Comprehensive Economic Partnership (RCEP) framework, which is already in force on January 1, 2022, involving countries such as Australia, Indonesia, and New Zealand.

Table 3. Bilateral trade costs (tariff equivalent) between Cambodia and its 30 major trading partners between 1993 and 2019 ($\sigma = 8$).

| Region | Trading Partner | $\tau_{1993}$ | $\tau_{2019}$ | Δ% |
|---|---|---|---|---|
| Southeast Asia | Thailand | 50.12 | 33.60 | −32.96 |
| | Vietnam | 78.81 | 40.59 | −48.49 |
| | Singapore | 67.78 | 54.11 | −20.16 |
| | Malaysia | 96.18 | 86.08 | −10.50 |
| | Indonesia | 145.43 | 114.14 | −21.51 |
| East Asia | Hong Kong (China)* | 53.52 | 1.35 | −97.48 |
| | Republic of Korea** | 201 | 56.82 | −71.73 |
| | Japan | 96.82 | 59.12 | −38.95 |
| | China | 197.06 | 62.22 | −68.43 |
| | Taiwan Province of China | 173.99 | 76.53 | −56.01 |
| South Asia | India | 264.06 | 167.32 | −36.63 |
| Oceania | Australia | 237.57 | 151.40 | −36.27 |
| | New Zealand | 308.38 | 167.00 | −45.85 |
| North America | United States | 196.59 | 98.70 | −49.79 |
| | Canada | 245.68 | 110.37 | −55.07 |
| Europe | United Kingdom | 255.07 | 66.60 | −73.89 |
| | Switzerland | 234.47 | 90.94 | −61.21 |
| | Belgium | 191.18 | 94.79 | −50.42 |
| | Denmark | 272.91 | 98.48 | −63.91 |
| | Netherlands | 130.32 | 110.30 | −15.36 |
| | France | 225.63 | 117.00 | −48.15 |
| | Germany | 147.90 | 121.92 | −17.57 |
| | Italy | 261.90 | 129.26 | −50.65 |
| | Spain | 433.46 | 138.05 | −68.15 |
| | Ireland | 431.88 | 144.74 | −66.49 |
| | Austria | 259.83 | 171.24 | −34.09 |
| | Sweden | 350.54 | 171.26 | −51.14 |
| | Finland | 198.40 | 172.69 | −12.96 |
| | Russian Federation | 189.14 | 196.23 | +3.74 |
| | Norway | 331.90 | 251.25 | −24.30 |

**Notes: All values are in percent and recorded with two decimal places—source**: Authors' calculation using Eq (8)

\* The bilateral trade costs between Cambodia and Hong Kong (China) have shown a negative value in some years. Therefore, the data for the year 2016 are used as the end-year data

\*\* For the Republic of Korea, trade costs are measured from the year 1995 due to the unavailability of data before the year 1995.

https://doi.org/10.1371/journal.pone.0311754.t003

Regarding the magnitudes of the changes, there have been significant shifts in the bilateral trade costs between Cambodia and its major trading partners. All trading partners have reduced their respective trade costs, except for the Russian Federation. It is clear from the findings that there is an opportunity for further trade optimization between Cambodia and the Russian Federation. Cambodia's bilateral trade costs with Hong Kong (China), the United Kingdom, the Republic of Korea, China, Spain, Ireland, Denmark, and Switzerland experienced a remarkable decline, exceeding 60% between 1993 and 2019. It is important to note that in 1993, Cambodia's bilateral trade costs with Spain and Ireland were extremely high, reaching 433.46% and 431.88%, respectively. However, by 2019, these figures dramatically decreased to 138.05% and 144.74%, showing a decline of over 65%. In contrast, Germany (–17.57%), the Netherlands (–15.36%), Finland (–12.96%), and Malaysia (–10.50%) had the lowest decline in trade costs.

**Table 4. Bilateral trade costs (tariff equivalent) between Cambodia and its ten largest trading partners between 1993 and 2019 ($\sigma = 8$).**

| Trading partner | $\tau_{1993}$ | $\tau_{2019}$ | $\Delta\%$ |
|---|---|---|---|
| China | 197.06 | 62.22 | −68.43 |
| United States | 196.59 | 98.70 | −49.79 |
| Thailand | 50.12 | 33.60 | −32.96 |
| Vietnam | 78.81 | 40.59 | −48.49 |
| Japan | 96.82 | 59.12 | −38.95 |
| Germany | 147.90 | 121.92 | −17.57 |
| United Kingdom | 255.07 | 66.60 | −73.89 |
| Canada | 245.68 | 110.37 | −55.07 |
| Republic of Korea* | 201 | 56.82 | −71.73 |
| Singapore | 67.78 | 54.11 | −20.16 |

Notes: All values are in percent and recorded with two decimal places—source: Authors' calculation using Eq (8).
* For the Republic of Korea, trade costs are measured from the year 1995 due to the unavailability of data before the year 1995.

https://doi.org/10.1371/journal.pone.0311754.t004

Table 4 displays the levels and the percentage reductions in Cambodia's relative bilateral trade costs with its ten largest trading partners between 1993 and 2019. The top ten largest trading partners based on the 2019 share of Cambodia's total trade volumes listed in descending order are China (23.93%), the United States (13.17%), Thailand (10.42%), Vietnam (8.58%), Japan (5.64%), Germany (3.35%), the United Kingdom (2.96%), Canada (2.67%), the Republic of Korea (2.47%), and Singapore (2.25%), according to the data from the IMF's DOTS database [72]. Concerning the percentage declines, only the United Kingdom, the Republic of Korea, and China reduced their bilateral trade costs with Cambodia by more than two-thirds between 1993 and 2019. In contrast, the lowest declines in trade costs were recorded in Thailand (−32.96%), Singapore (−20.16%), and Germany (−17.57%).

Regarding its neighboring countries, Cambodia's bilateral trade costs were 33.6% with Thailand and 40.59% with Vietnam in 2019. These figures show that Thailand and Vietnam have the second and third lowest trade costs, respectively, following Hong Kong (China) (see Table 3). In 2019, Thailand and Vietnam were Cambodia's third and fourth-largest trading partners, respectively, behind China and the United States. Notably, Thailand and Vietnam are the only two of Cambodia's top 10 trading partners with trade costs below 50%. There are several reasons for low trade costs with these countries. First, Cambodia's geographic advantage as a neighboring country of Thailand and Vietnam allows it to enjoy lower transportation costs when trading with these countries. Next, Cambodia and its neighbors have developed close economic ties due to their long-standing shared history and cultural similarities. Finally, they have been a part of the ASEAN Free Trade Area (AFTA) since 1999, which is significant for reducing trade barriers among its members. This finding aligns with the literature showing that trade integration has rapidly accelerated across geographically close countries. For example, Novy [24] found that in the year 1993, the bilateral trade costs (tariff equivalent) between the two neighbors (the United States and Canada) were 31%, which is significantly lower than trade costs between Cambodia and Thailand (50.12%) and Cambodia and Vietnam (78.81%) in the same year. These numbers show that in 1993, Cambodia's bilateral trade costs with Thailand and Vietnam were 61.68% and 154.23%, respectively, higher than those between the United States and Canada. In 2019, the bilateral trade costs between Cambodia and Thailand stood at 33.60%. This figure is comparable to 31% of the United States and Canada bilateral trade costs in 1993 and 33% of the United States and Mexico bilateral trade costs in 2000, as

Table 5. Cambodia's average trade costs from 1993 to 2019 ($\sigma = 8$).

| Year | 1993 | 1994 | 1995 | 1996 | 1997 | 1998 | 1999 | 2000 | 2001 |
|---|---|---|---|---|---|---|---|---|---|
| Trade Costs | 139.7 | 141.76 | 144.46 | 150.45 | 140 | 136.51 | 139 | 132.82 | 133.18 |
| **Year** | **2002** | **2003** | **2004** | **2005** | **2006** | **2007** | **2008** | **2009** | **2010** |
| Trade Costs | 127.57 | 130.16 | 126.97 | 131.87 | 131.29 | 131.97 | 132.17 | 124.22 | 118.94 |
| **Year** | **2011** | **2012** | **2013** | **2014** | **2015** | **2016** | **2017** | **2018** | **2019** |
| Trade Costs | 120.09 | 117.38 | 111.57 | 119.73 | 109.97 | 103.27 | 99.28 | 96.55 | 90.21 |

**Notes**: All values are in percent and recorded with two decimal places—*source*: Authors' calculation using Eq (10).

https://doi.org/10.1371/journal.pone.0311754.t005

reported in Novy's survey [24]. Although there have been significant declines in the bilateral trade costs between Cambodia and its neighboring countries over the past two decades, these costs have remained higher than those observed between highly developed countries nearly 30 years ago. This comparison highlights the significant progress made by highly developed countries in reducing trade costs over time.

Another noteworthy finding is that Germany and Canada, two of Cambodia's top ten trading partners, had trade costs exceeding 100% in 2019 (refer to Table 4). Simultaneously, the bilateral trade costs between Cambodia and the United States were also relatively high, approaching 100%. These three countries' trade costs were higher than the average of 90.21% in 2019 (see Table 5).

### 5.2 Measurement results and analyses of Cambodia's average trade costs

For a comprehensive understanding of Cambodia's trade costs and their drastic changes over time, it is crucial to measure the country's average trade costs with all of its trading partners. However, due to the limited data availability, this study includes only 30 of Cambodia's major trading partners. In this sub-section, we compute Cambodia's average trade costs using a model-based equation. This method is considered superior to the simple or trade-weighted average, as it does not give much weight to large trading partners [27]. After calculating the bilateral trade costs between Cambodia and each trading partner using Eq (8), we compute Cambodia's average trade costs based on Eq (10), and the estimated results are shown in Table 5.

Table 5 shows that Cambodia's average trade costs (ad valorem equivalent) peaked at 150.45% in 1996. We can observe that before the country joined the ASEAN in 1999, its average trade costs remained high, exceeding 136% over the years from 1993 to 1999. Following Cambodia's accession to the ASEAN, there has been a notable reduction in its average trade costs. Specifically, its average trade costs decreased from 139% in 1999 to 132.82% in 2000, equivalent to a decline of 4.44%. Despite becoming a member of the WTO in 2004, Cambodia had experienced persistently high and volatile trade costs until 2014. Since then, these costs have gradually decreased, falling below 100% in 2017 and reaching 90.21% in 2019.

As displayed in Fig 3, Cambodia's average trade costs have shown a clear downward trend since 2014. Between 1993 and 2019, Cambodia's average trade costs decreased by 35.43%. These costs fell by a huge margin of 24.66% between 2014 and 2019 alone, adding to a considerable decline in the overall trade costs between 1993 and 2019. These findings suggest that Cambodia has committed to fulfilling its trade obligations under the WTO framework and has consistently increased its trade openness to global markets.

In order to gain insights into the scales and the evolutions of Cambodia's trade costs over the past two decades, it is crucial to conduct a comparative analysis of its average trade costs with other countries, as documented in the literature that used the same measuring methods.

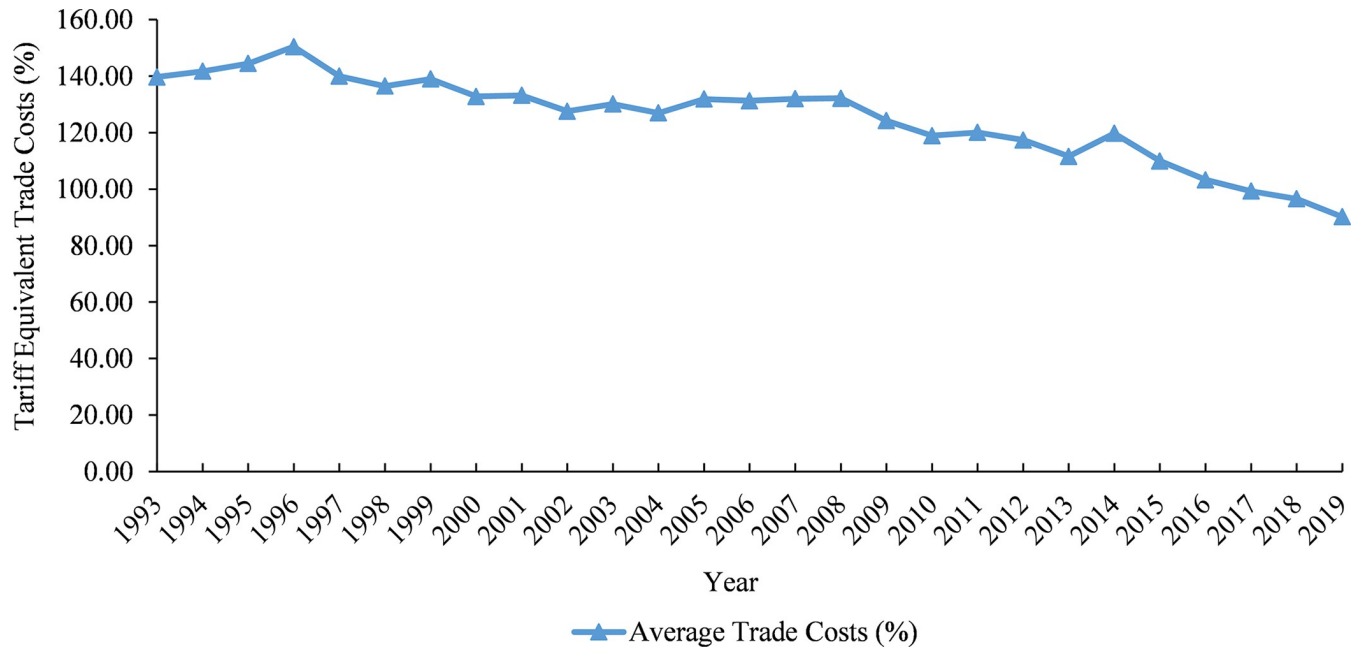

**Fig 3. The trend of Cambodia's average trade costs from 1993 to 2019 (*σ* = 8). Source:** Authors' calculation using Eq (10).

https://doi.org/10.1371/journal.pone.0311754.g003

In the year 2000, Cambodia's average trade costs (tariff equivalent) stood at 132.82% (see Table 5). This number is 41.29% higher than the average trade costs of 13 OECD countries, which were recorded at 94% in the same year, as reported by Novy [24]. In 2004, the average trade costs of 37 major developed and industrializing countries were only 66%, according to a study conducted by Milner and Mcgowan [31]. This figure is almost half as low as the average trade costs of Cambodia, which were nearly 127% in the same year. In 2019, Cambodia's average trade costs were approximately 90%. It is evident that although there has been a downward trend in Cambodia's average trade costs in recent years, they remain 36% higher than those in major developed and industrializing countries 16 years ago.

As shown in Fig 3, Cambodia's average trade costs have declined more rapidly since 2014. We provide a few possible explanations for this phenomenon. First, most of Cambodia's top trading partners recovered from major crises, including the 2008–2009 global financial crisis, the European debt crisis in 2010, and the European Union recession in 2013. Therefore, various countries and regions relaxed trade protectionism, which led to a decline in trade costs between Cambodia and its major trading partners. Second, trade costs between Cambodia and its major trading partners along the Belt and Road Initiative (BRI) corridors, especially China, have decreased significantly since 2014, contributing to the overall decline in Cambodia's average trade costs. This can be attributed to infrastructure development, tariff rate adjustments, and other measures implemented as part of the BRI, launched by President Xi Jinping in September 2013. Thus, to delve deeper into the impact of the BRI on Cambodia's trade costs, as discussed in the next section, we classify 30 trading partners into two groups: the BRI trading partners and the non-BRI trading partners.

Finally, the rapid decline in average trade costs after 2014 coincided with Cambodia's transformation from a low-income to a lower-middle-income country in 2015, according to the World Bank's country classifications by income level [41]. We can observe that Cambodia experienced consistently high trade costs throughout its status as a low-income country from 1993 to 2014. This finding aligns with the literature indicating that trade costs decrease as the country's per capita income increases. Low-income countries have not witnessed any significant improvement in their trade costs compared to other income groups, and those countries continue to face difficulties in reducing trade barriers [27]. Therefore, to provide a more in-depth understanding of the dynamic changes in Cambodia's average trade costs with major trading partners based on the different levels of their economic development, we classify 30 trading partners into two groups: developed economies and developing & emerging economies. With the same measure stated in Eq (10), Cambodia's average trade costs with its trading partners under different classifications can be determined, and the results are elaborated in the following sections.

**5.2.1 A comparison between Cambodia's trade costs with developing economies and developed economies.** To gain a comprehensive understanding of various degrees of trade costs among trading partners based on their respective levels of economic development, we use the economy groupings classified by the International Monetary Fund (IMF) to divide the whole sample into two groups: developed economies and developing & emerging economies [76]. Seven developing and emerging economies include China, India, Indonesia, Malaysia, the Russian Federation, Thailand, and Vietnam. Twenty-three developed economies include Australia, Austria, Belgium, Canada, Denmark, Finland, France, Germany, Hong Kong (China), Ireland, Italy, Japan, the Republic of Korea, the Netherlands, New Zealand, Norway, Singapore, Spain, Sweden, Switzerland, the Taiwan Province of China, the United Kingdom, and the United States.

In terms of the scales, it can be seen from Fig 4 that the average trade costs between Cambodia and developing and emerging economies were consistently lower than those of the developed economies from 1993 to 2006. From 2007 to 2012, average trade costs between Cambodia and the two groups were similar. However, Cambodia's average trade costs with developing and emerging economies experienced a slight rebound in 2009 and 2010, surpassing the average trade costs of developed economies by 11.25% and 3%, respectively. Between

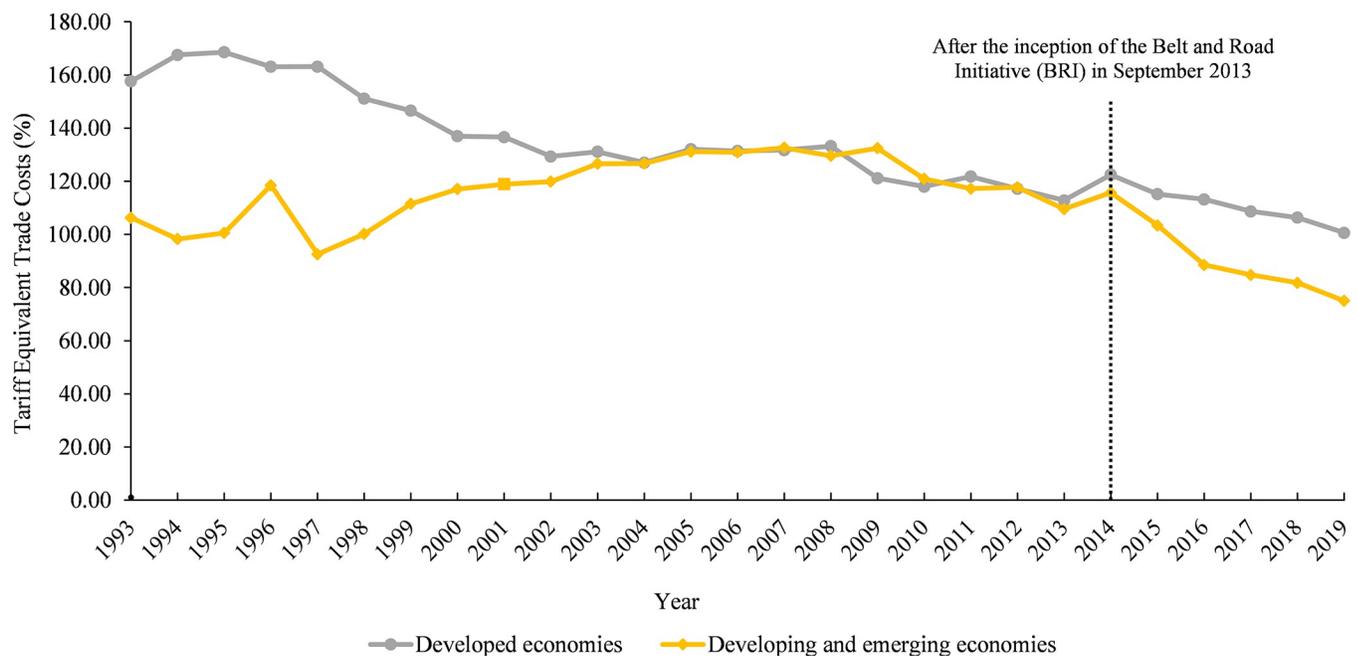

**Fig 4. Changes in Cambodia's average trade costs with developed economies and developing & emerging economies from 1993 to 2019 ($\sigma$ = 8). Source**: Authors' calculation using Eq (10).

https://doi.org/10.1371/journal.pone.0311754.g004

2013 and 2014, there was a marginal rise in average trade costs between Cambodia and the two groups. One potential explanation for this phenomenon might be attributed to the 2008–2009 global financial crisis and the subsequent European Union recession in 2013. These economic downturns contributed to a rise in trade protectionism across different countries and regions, causing an increase in trade costs. These findings are consistent with the literature, suggesting that an uncertainty shock during a crisis directly hinders cross-border trade flows while simultaneously increasing average trade costs for all trading partners [77].

Contrary to what one would expect, Cambodia's average trade costs with developing and emerging economies are lower than those with developed economies when engaging in trade. This finding also aligns with the substantial contribution of developing and emerging economies to Cambodia's total trade volumes. For instance, seven emerging and developing economies, namely China, India, Indonesia, Malaysia, the Russian Federation, Thailand, and Vietnam, made up almost 48% of Cambodia's overall trade in 2019. In contrast, the remaining twenty-three developed economies accounted for only 46.4% of the country's total trade. Due to its proximity, Cambodia has benefited from lower transportation costs when trading with developing and emerging economies like Thailand, Vietnam, China, India, Indonesia, and Malaysia. Therefore, their trade costs are less likely to be affected by their level of per capita income.

Regarding the magnitudes of the changes, the average trade costs of developed economies showed a downward trend after reaching a peak at 168.57% in 1995. These costs decreased by 36.17%, from 157.61% in 1993 to 100.60% in 2019. In contrast, the average trade costs of developing and emerging economies increased from 106.29% in 1993 to a peak of 132.67% in 2007. Between 1993 and 2019, average trade costs between Cambodia and developing and emerging economies decreased by 29.42%, from 106.29% in 1993 to 75% in 2019. The findings show that since 2014, there has been a notable decline in the trend of average trade costs between Cambodia and both groups. Cambodia's average trade costs with developing and emerging economies decreased by 35.17%, from 115.71% in 2014 to 75% in 2019, equivalent to 8.3% per year. This figure surpasses the 17.85% decline in developed economies between 2014 and 2019, where average trade costs dropped from 122.46% in 2014 to 100.60% in 2019, corresponding to 3.86% annually. The decline in trade costs is likely due to developing and emerging countries accelerating the liberalization of their trade policies. In addition, contributing to this reduction in trade costs is also influenced by Cambodia's trade openness and economic integration with developing and emerging countries through various regional trade agreements (RTAs) under the framework of the ASEAN Free Trade Area (AFTA effective from 1999), the ASEAN-China Free Trade Area (ACFTA effective from 2005), and the ASEAN-India Free Trade Area (AIFTA effective from 2010). In this analysis, all trading partners categorized as developing and emerging economies comprise lower-middle-income countries (India and Vietnam) and upper-middle-income countries (China, Indonesia, Malaysia, the Russian Federation, and Thailand) according to the World Bank country classifications by income levels in 2019 [78]. Therefore, this finding is consistent with the existing literature indicating that developing countries in the upper-middle-income bracket have achieved a more accelerated reduction in trade costs compared to developed countries [27]. It also reflects that trade frictions remain despite long-standing trade ties between Cambodia and developed economies. For example, in May 2015, Cambodian bicycle exports to the European Union market were subject to a 48.5% import duty under the rule of origin [79].

**5.2.2 A comparison between Cambodia's trade costs with the Belt and Road Initiative (BRI) and the non-BRI trading partners.** In light of the growing interest in the Belt and Road Initiative (BRI) among academics and policymakers in recent years, we also consider the impact of the BRI on Cambodia's trade costs. The year 2023 marked the tenth anniversary of

the BRI since it was introduced by President Xi Jinping in September 2013. By the end of August 2023, more than 150 countries and over 30 international organizations have participated and signed cooperation agreements with China under the BRI framework [80]. Cambodia was among the first countries to support the BRI projects, and the Royal Government of Cambodia (RGC) signed the Memorandum of Understanding (MoU) with China in 2016 under the BRI's framework to develop the infrastructure [81, 82]. The existing literature on trade, investment, and infrastructure development related to the BRI has demonstrated that countries or regions adjacent to China have experienced the extensive scales of development brought about by this initiative [36, 83, 84]. Hence, there could be an expansion of trade between Cambodia and its trading partners along the BRI corridors despite Cambodia not being a signatory to the BRI with other countries besides China. We consult several sources to determine the major trading partners along the BRI corridors. First, the initial list of 64 countries along the BRI corridors and China was published by the Chinese government in 2015, including Cambodia, China, India, Indonesia, Malaysia, the Russian Federation, Singapore, Thailand, and Vietnam [82]. We also include Hong Kong (China), Japan, the Republic of Korea, and the Taiwan Province of China as Cambodia's trading partners along the BRI corridors, according to the lists identified by the OECD and the World Bank [36, 85]. Thus, we finally classify 12 major trading partners of Cambodia along the BRI corridors for our analysis, which comprise China, Hong Kong (China), India, Indonesia, Japan, the Republic of Korea, Malaysia, the Russian Federation, Singapore, the Taiwan Province of China, Thailand, and Vietnam. We also categorize the other 18 major trading partners of Cambodia as the non-BRI countries, which include Australia, Austria, Belgium, Canada, Denmark, Finland, France, Germany, Ireland, Italy, the Netherlands, New Zealand, Norway, Spain, Sweden, Switzerland, the United Kingdom, and the United States.

Fig 5 indicates that average trade costs between Cambodia and its trading partners along the BRI corridors decreased by nearly 34.78%, from 108.16% in 2014 to 70.54% in 2019, equivalent to 8.2% per year. This number is significantly higher than the 16.14% decline in the non-BRI trading partners between 2014 and 2019, with the average trade costs falling from 136.34% in 2014 to 114.34% in 2019, corresponding to 3.46% per year. Overall, average trade costs between Cambodia and the trading partners along the BRI corridors have decreased by more than twice as fast as the non-BRI trading partners since 2014. A possible explanation for the rapid decline in trade costs between Cambodia and its trading partners along the BRI corridors may also reflect better infrastructure development and improved trade facilitation. Furthermore, the large-scale BRI development projects are more likely to have significant impacts on the markets of infrastructure-construction-related products, such as construction equipment, cement, iron, and steel, among others, between Cambodia and its trading partners along the BRI corridors, especially those along the China-Indochina Peninsula Economic Corridor. This finding is consistent with the literature. For example, a study conducted by de Soyres et al. [36] demonstrates that economies located along the BRI corridors, where transportation infrastructure projects have been developed, witness significant benefits. These benefits include a notable reduction in shipment time of up to 11.9% and a decrease in trade costs by up to 10.2%. In comparison, the global average decline in shipment time ranges from 1.2% to 2.5%, leading to a decrease in the global aggregate trade costs between 1.1% and 2.2%.

### 5.3 Decomposing the growth of Cambodia's trade

Cambodia's adoption of a market-based economy in the early 1990s, followed by its membership in the ASEAN in 1999 and the WTO in 2004, has contributed significantly to the rapid growth of its trade. From 1993 to 2019, Cambodia's total trade volumes increased 46.5 times,

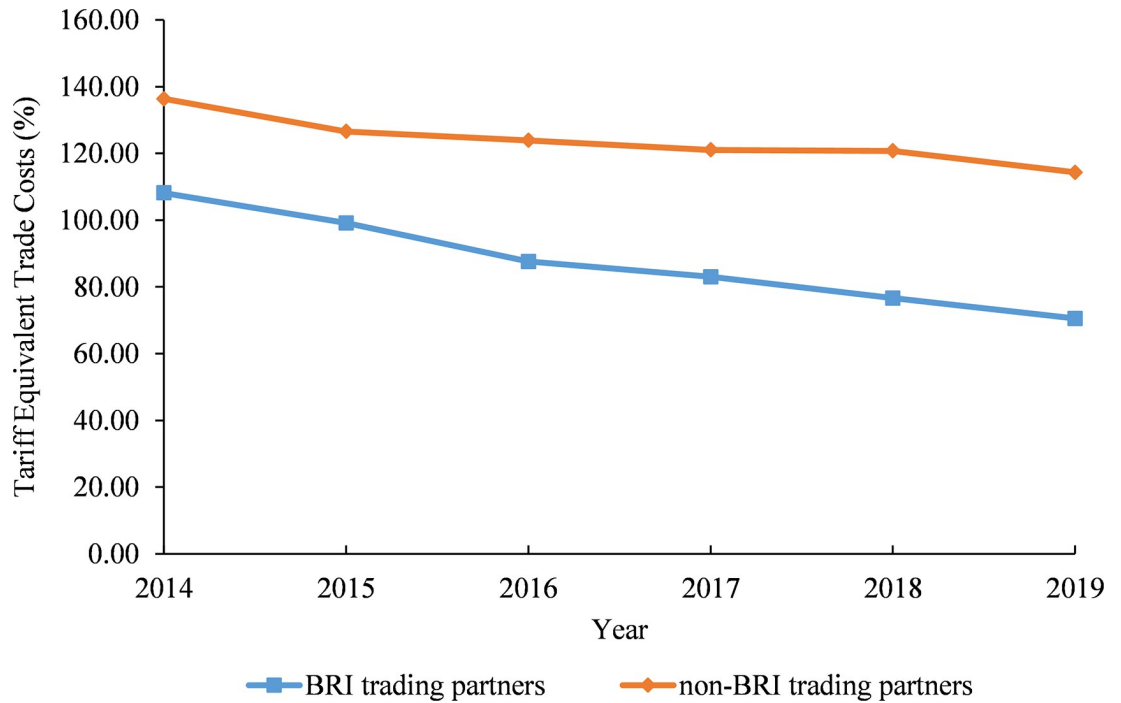

**Fig 5. Changes in Cambodia's average trade costs with the Belt and Road Initiative (BRI) and the non-BRI trading partners from 2014 to 2019 ($\sigma = 8$). Source:** Authors' calculation using Eq (10).

https://doi.org/10.1371/journal.pone.0311754.g005

from US$ 755 million to US$ 35.1 billion. Therefore, a couple of questions arise: What factors have led to the growth of Cambodia's trade? How do trade costs contribute to trade expansion? This section decomposes Cambodia's trade growth between 1993 and 2019 into three components: income, trade costs, and multilateral resistance. This portion helps us understand the determinants that have driven the growth of Cambodia's trade over the past two decades.

**5.3.1 Decomposition of bilateral trade growth between Cambodia and its trading partners by region.** Table 6 presents the decomposition of the Cambodian trade expansion with its top 30 trading partners across the six regions between 1993 and 2019. The analysis focuses on three key factors contributing to this growth: income, trade costs, and multilateral resistance. We organize the region in descending order based on its relative trade volumes with

**Table 6. Decomposition of bilateral trade growth between Cambodia and its trading partners by region between 1993 and 2019.**

| Region | Due to GDP growth (%) | Due to the decline in trade costs (%) | Due to the decline in multilateral resistance (%) | Total (%) |
|---|---|---|---|---|
| Southeast Asia | +113.89 | +27.30 | −41.18 | = 100 |
| East Asia | +64.52 | +82.73 | −47.25 | = 100 |
| North America | +40.16 | +43.48 | +16.37 | = 100 |
| Europe | +64.22 | +52.70 | −16.92 | = 100 |
| South Asia | +92.63 | +24.38 | −17.01 | = 100 |
| Oceania | +62.15 | +45.86 | −8.01 | = 100 |
| Average | +59.65 | +56.69 | −16.34 | = 100 |

**Notes**: The sum of the GDPs of each pair in 2019 is used as a weight to calculate the group's average in each region and the whole sample. The base-year data are computed as the three-year average from 1993 to 1995 to avoid annual volatility during Cambodia's economic reforms to a market-oriented economy in 1993 and to establish a balanced sample for all 30 trading partners—**source:** Authors' calculation using Eq (16).

https://doi.org/10.1371/journal.pone.0311754.t006

Table 7. Decomposition of bilateral trade growth between Cambodia and its ten largest trading partners between 1993 and 2019.

| Trading Partner | Growth in trade (%) | Due to GDP growth (%) | Due to the decline in trade costs (%) | Due to the decline in multilateral resistance (%) | Total (%) |
| --- | --- | --- | --- | --- | --- |
| China | 1,145.36 | +73.87 | +65.55 | −39.42 | = 100 |
| United States | 1,099.13 | +39.95 | +40.72 | +19.33 | = 100 |
| Thailand | 477.28 | +101.23 | +33.48 | −34.71 | = 100 |
| Vietnam | 708.06 | +108.76 | +44.30 | −53.06 | = 100 |
| Japan | 578.07 | +39.22 | +121.64 | −60.86 | = 100 |
| Germany | 736.74 | +44.66 | +48.95 | +6.39 | = 100 |
| United Kingdom | 794.18 | +49.13 | +119.1 | −68.22 | = 100 |
| Canada | 1030.6 | +42.72 | +76.84 | −19.56 | = 100 |
| Republic of Korea | 741.68 | +63.73 | +115.71 | −79.44 | = 100 |
| Singapore | 394.18 | +139 | +26.55 | −65.54 | = 100 |

**Notes:** The base-year data are computed as the three-year average from 1993 to 1995 to avoid annual volatility during Cambodia's economic reforms to a market-oriented economy in 1993 and to establish a balanced sample for all 30 trading partners—**source:** Authors' calculation using Eq (16).

https://doi.org/10.1371/journal.pone.0311754.t007

Cambodia as of 2019. The finding illustrates that the growth of the bilateral trade between Cambodia and its major trading partners has been attributed mainly to the rise in income and the decline in trade costs. The average contribution rate of the income growth to the bilateral trade growth is +59.65%, which is slightly higher than the contribution of the decline in trade costs (+56.69%). With an average contribution rate of −16.34%, the decrease in multilateral resistance impedes the growth of the Cambodian trade.

Because of the expanding overall scale of the GDP, there have been significant increases in the bilateral trade between Cambodia and the six regions: Southeast Asia, East Asia, South Asia, Oceania, Europe, and North America. Among these regions, the growth of the Cambodian trade with Southeast Asia and South Asia has been driven mainly by the rise in income, which accounts for over 90% of the total. The contribution from the GDP growth of the other three regions, including East Asia, Europe, and Oceania, is also substantial, exceeding 60%. Although the total income growth has a lower influence on trade growth in North America than in other regions, it still amounts to 40.16%.

These geographic differences reflect that the economic growth of Southeast Asian countries has been the most crucial factor in the expansion of trade with Cambodia, with individual countries' contributions ranging from 101.23% in Thailand to 139% in Singapore (see Table 7). Besides countries in Southeast Asia, the bilateral trade growth of Cambodia has also been attributed to the rise of the GDP of China (73.87% from Table 7) in East Asia (64.52%) and India in South Asia (92.63%). Southeast Asia's growing economies are primarily responsible for the increase in income that has fueled the expansion of their bilateral trade with Cambodia, followed by South Asia and East Asia. Since the GDP growth of developing countries in these regions outperforms most developed nations, there is great potential for expanding Cambodian trade in the future. This finding is consistent with previous studies showing the significance of economic growth in enhancing trade between countries [24, 62].

We can observe that the decline in trade costs has a more substantial influence on the expansion of Cambodia's trade with East Asia and North America than income growth. Among the six regions, the largest impact on promoting trade growth between Cambodia and East Asia trading partners comes from the decrease in trade costs, which accounts for 82.73%. The Cambodian trade volumes with Europe, North America, and Oceania have also increased by more than 40% due to the reduction in trade costs.

Overall, the decline in multilateral resistance has reduced the growth of the Cambodian trade. The average contribution rate of the decrease in multilateral resistance in the sample period is –16.34%. Compared to other regions, Cambodia's trade with Southeast Asia and East Asia is negatively impacted by the decrease in multilateral resistance, with the contribution rate reaching –40%. This finding reflects the ability of trading partners in Southeast Asia and East Asia to reduce trade costs with their trading partners. As a result, Cambodia may experience the consequences of the trade diversion, in which its trading partners choose to trade more with countries in Southeast Asia and East Asia rather than with Cambodia. For example, among the six regions, the decline in trade costs contributes the most to the bilateral trade growth between Cambodia and East Asia (+82.73%). However, these gains have been counterbalanced by the negative impact of a decline in multilateral resistance, bringing the overall contribution rate of a decrease in trade costs to +35.48% (+82.73%–47.25%). In contrast, the rise of North American multilateral resistance has promoted Cambodia's foreign trade.

**5.3.2 Decomposition of bilateral trade growth between Cambodia and its ten largest trading partners.** Table 7 decomposes the growth of the bilateral trade between Cambodia and its ten largest trading partners into three main factors: income, trade costs, and multilateral resistance. As of 2019, Cambodia's top ten trading partners, arranged in descending order, are China, the United States, Thailand, Vietnam, Japan, Germany, the United Kingdom, Canada, the Republic of Korea, and Singapore.

It is worth mentioning that the share of Cambodia's total trade volumes with each country exceeded two percent, and these ten countries accounted for almost 76% of Cambodia's total trade volumes with the world as of 2019, according to the IMF's DOTS database [72]. Over the period from 1993 to 2019, we can observe that the growth of the bilateral trade between Cambodia and China is the largest, expanding by 1,145.36%. Cambodia has also increased its bilateral trade with the United States and Canada by more than 1,000% among the top ten trading partners. It is important to note that China is Cambodia's largest trading partner and primary import market, while the United States is its major export market. As of 2019, China accounted for almost 24% of Cambodia's total trade volumes, followed by the United States at about 13%. The trade volumes of these two countries were approximately equivalent to the combined trade volumes of the eight remaining top 10 trading partners [72]. With its neighbors, Cambodia has experienced trade growth of 477.28% with Thailand and 708% with Vietnam. Cambodia's trade has grown rapidly, with the top five largest economies in the world as of 2019 (the United States, China, Japan, Germany, and the United Kingdom), all of which have climbed by more than 500%.

It is evident that the primary factors driving Cambodia's bilateral trade growth are the rise in income and the decline in trade costs. From the perspective of the income contribution, the trade growth between Cambodia and Singapore is mainly due to income growth (139%). In comparison, the growth of trade between Cambodia and Japan because of income growth is just 39.22%. We can observe that the expansion of Cambodian bilateral trade with China, Singapore, Thailand, and Vietnam is primarily driven by the rise in income, which accounts for over 70% of the growth.

The contribution of the decline in trade costs is the highest between Cambodia and Japan, reaching 121.64%, followed by the United Kingdom (+119.1%) and the Republic of Korea (+-115.71%). The increase in Cambodia's trade with Canada, Germany, Japan, the Republic of Korea, the United Kingdom, and the United States is mainly explained by the decline in bilateral trade costs. Overall, the contribution rate of the decline in bilateral trade costs between Cambodia and other countries except for Japan, the United Kingdom, the Republic of Korea, Canada, and China is lower than 50%, ranging from +26.55% in Singapore to +48.95% in Germany.

Apart from the United States and Germany, Cambodia's bilateral trade with other major trading partners is diverted due to the decrease in multilateral resistance. The profound impact of multilateral resistance is seen in the Republic of Korea (–79.44%), followed by the United Kingdom (–68.22%), Singapore (–65.54%), Japan (–60.86%) and Vietnam (–53.06%). This finding suggests that there has been a substantial decrease in multilateral trade barriers between these countries and their trading partners, resulting in a significant trade diversion impact on Cambodia. For example, in the case of Vietnam, the decline in multilateral resistance led to a decrease in trade between Cambodia and Vietnam by –53.06%. Vietnam has made substantial progress in its trade openness with the world since its membership in the WTO in 2007. Vietnam's rapid elimination of trade barriers with its trading partners has also diverted trade away from Cambodia. Though the decline in the bilateral trade costs between Cambodia and Vietnam significantly boosts the bilateral trade between the two countries by +- 44.30%, the adverse impact of multilateral resistance diminishes this positive driving force to –8.76% (+44.30%–53.06%). Overall, the expansion of the bilateral trade between Cambodia and Vietnam has been driven mainly by income growth. Therefore, there will be the great potential for trade growth in the future as the two countries increase their living standards. In contrast, the impact of multilateral resistance turns positive for Germany and the United States, indicating that the increase in multilateral resistance will make Cambodia's trade attractive for these two countries.

## 6. Conclusions

### 6.1 Major findings

This study employs Novy's micro-founded method of trade costs to measure Cambodia's trade costs with its top 30 trading partners from 1993 to 2019. The study also decomposes the growth of the Cambodian trade during the same period into three driving forces: income, trade costs, and multilateral resistance. After the in-depth analyses, the following conclusions have been drawn:

Between 1993 and 2019, Cambodia's average trade costs decreased significantly by 35.43 percent, from 139.7 percent to 90.2 percent. There were fluctuations in its average trade costs until 2014, even though Cambodia joined the WTO in 2004. The remarkable 24.66 percent decline between 2014 and 2019 alone was primarily responsible for the overall decrease in the country's average trade costs between 1993 and 2019. Regarding the individual trading partners, Cambodia's bilateral trade costs with Hong Kong (China) and its neighboring countries (Thailand and Vietnam) have dropped to less than 50 percent. The bilateral trade costs of these three trading partners, along with Singapore, remained below 100 percent over the years from 1993 to 2019, which were consistently lower than those of other trading partners. In terms of the regions, Cambodia has enjoyed lower trade costs with its major trading partners in Southeast Asia and East Asia than with those in South Asia, Oceania, Europe, and North America.

As far as the magnitudes of the changes are concerned, all trading partners have reduced their trade costs, except for the Russian Federation. Thus, there is an opportunity for further optimization of trade relations between Cambodia and the Russian Federation if both countries can decrease their bilateral trade costs. Notably, China, Denmark, Hong Kong (China), Ireland, the Republic of Korea, Spain, Switzerland, and the United Kingdom experienced a remarkable decline in bilateral trade costs with Cambodia, exceeding 60 percent between 1993 and 2019. On the other hand, Finland, Germany, Malaysia, and the Netherlands all saw a decrease in trade costs of less than 20 percent. Among the top ten trading partners, Cambodia's bilateral trade costs with China, the United States, Thailand, Vietnam, Japan, the United

Kingdom, the Republic of Korea, and Singapore have declined to less than 100 percent. In contrast, Cambodia's bilateral trade costs with Germany and Canada have remained high, surpassing 100 percent. Furthermore, only Canada, Germany, and the United States had bilateral trade costs higher than the average among the top ten trading partners in 2019. Therefore, more attention is needed to enhance trade facilitation and improve the trade structure among these three trading partners.

In terms of trade costs among different groups, Cambodia's average trade costs with developing and emerging economies are lower than those with developed economies. Since 2014, Cambodia's average trade costs with developing and emerging economies have declined at double the rate of those with developed economies. Between 2014 and 2019, average trade costs between Cambodia and the trading partners along the Belt and Road Initiative (BRI) corridors decreased by 34.78 percent, twice as fast as observed with the non-BRI trading partners. These findings demonstrate that improvements in trade facilitation have led to a gradual decline in Cambodia's average trade costs, primarily attributed to the increasing investment in infrastructure projects in Cambodia and its major trading partners along the BRI corridors.

Regarding the decomposition analysis of trade growth, we find that the expansion of Cambodia's trade between 1993 and 2019 was driven by the rise in income (59.65 percent), the decline in trade costs (56.69 percent), and the decline in multilateral resistance (–16.34 percent). The expansion of trade opportunities among Cambodia's major trading partners and global markets led to a decline in their respective multilateral resistance, which diverted Cambodia's trade growth by more than 16 percent.

## 6.2 Policy implications and recommendations

This study will contribute to a better understanding of Cambodia's development process and endeavors toward its trade integration with the world over the past two decades, starting from the initial stages of its economic reforms in the early 1990s until the recent period. Moreover, these findings have practical implications in providing new insights for formulating trade strategies to support Cambodia's ambition to transition from a least-developed country to an upper-middle-income country by 2030 and eventually progress towards a high-income country by 2050.

Our findings suggest that Cambodia has the potential to optimize its trade growth by focusing on its economic diplomacy strategies with the trading partners exhibiting high economic growth potential and those showing substantial reductions in trade costs. Over the past two decades, Cambodia has made considerable progress in its integration into the global economy through its membership in the ASEAN in 1999 and the WTO in 2004. As a result, the country's overall trade costs have been decreasing steadily. However, these costs have remained higher than those in major developed and newly industrialized countries. First, Cambodia should explore the possibility of reducing its bilateral trade costs among its major trading partners, especially those countries that have experienced high trade costs. For instance, major emerging economies, such as India, Indonesia, and the Russian Federation, demonstrate huge economic potential. However, their bilateral trade costs with Cambodia have remained high, thereby hindering the potential for trade expansion. Thus, it is essential to focus on ongoing efforts to mitigate trade frictions and improve trade facilitation among these countries.

Second, Cambodia should reduce its bilateral trade costs with major import trading partners that supply the necessary goods for the country's production, such as intermediate goods, raw materials, and capital goods. To achieve these goals, the government can implement a series of measures. (i) Cambodia should offer tariff exemptions on intermediate goods, raw materials, and capital goods, thus incentivizing efficient production processes. (ii) The government should continue to adopt the opening-up policies that foster international trade and economic cooperation. These strategies can be achieved by seeking the opportunity to sign more regional trade agreements (RTAs) or economic partnership agreements with its major trading partners. The aim is to promote mutually beneficial trade cooperation and enhance the country's integration into global value and supply chains. (iii) The government should prioritize the investment in infrastructure development. The effective investment strategy should involve establishing economic zones, free-trade zones, free-trade ports, industrial parks, and logistic hubs near borders and ports. Implementing this measure is significant in facilitating the development of industrial clusters that bring together diverse enterprises, thereby minimizing trade costs associated with shipping goods. (iv) The government should eliminate the different obstacles created by various institutions for transporting goods across borders, which are imperative for reducing trade costs.

In light of the current backlash against globalization, Cambodia should advocate for more openness for international trade and support regional and global efforts to promote trade facilitation between countries or regions.

### 6.3 Limitations of the study

Despite providing significant policy insights into the scales, evolutions, and impacts of Cambodian trade costs over the past two decades, this study has certain caveats. First, this paper only examines Cambodia's trade costs with its top 30 trading partners. Measuring Cambodia's average trade costs with all its trading partners is necessary to comprehensively understand the evolutions of the country's aggregate trade barriers over time. However, the limited data availability makes it impossible to determine Cambodia's aggregate trade costs with all its trading partners. Second, this research does not cover the most recent economic partnership agreements and regional trade agreements (RTAs), including the Cambodia-Hong Kong (China) Free Trade Agreement in February 2021 under the ASEAN-Hong Kong (China) Free Trade Agreement (AHKFTA), the Cambodia-China Free Trade Agreement (CCFTA) in January 2022, the Regional Comprehensive Economic Partnership (RCEP) in January 2022, the Cambodia-Republic of Korea Free Trade Agreement (CKFTA) in December 2022, and the Cambodia-United Arab Emirates Comprehensive Economic Partnership Agreement (CAM-UAE CEPA) in 2023, which are already in force. With the implementation of these agreements, Cambodia can anticipate a significant reduction in its average trade costs.

Furthermore, our analysis focuses solely on total trade costs in the goods sector and does not account for disaggregated trade costs within different industries. Finally, it is important to acknowledge the limitations inherent in the modeling approaches, which are based on several assumptions. As a result, the actual trade costs that Cambodia has experienced in practice may differ from the estimates generated by the models. Therefore, the findings should be interpreted with caution.

### 6.4 Future research directions

This work can serve as a benchmark for future studies, as it is the first research endeavor to measure Cambodian trade costs. Several areas require further investigation. (i) Expanding the sample size will provide a more thorough understanding of trade costs from a diverse range of countries (both developed and developing countries) in different regions. This will contribute to formulating trade policies that are well-suited to the specific circumstances of countries, considering the disparities in trade costs among countries and regions. (ii) In recent years, the Royal Government of Cambodia (RGC) has made efforts to sign, negotiate, and propose multiple RTAs and economic partnership agreements with different trading partners. In addition, there has been a significant amount of investment in infrastructure, aiming to reduce trade costs. For example, Cambodia recently commenced the groundbreaking construction of the "Funan Techo Canal", spanning over 180 kilometers, on August 5, 2024. Given that this mega-project is imperative to Cambodia's waterway transport infrastructure, its completion within five years will significantly reduce the country's overall trade costs. Therefore, the determinants influencing trade costs, such as the current RTAs and infrastructure development, should also be explored. (iii) For a better understanding of Cambodia's participation in global value chains (GVCs), measuring trade costs using trade in value-added should be considered in future research. (iv) Investigating trade costs associated with various categories of goods or industries with high potential for growth, including agriculture, manufacturing, and services, requires attention in the future. This could yield findings that would be valuable for researchers, scholars, and policymakers.

## Supporting information

**S1 Appendix. Details of the data descriptions.**
(DOCX)

## Acknowledgments

The authors would like to express their gratitude to the editors and the four anonymous reviewers for their critical and constructive comments as well as insightful suggestions, which significantly contributed to improving the quality of the earlier version of the manuscript.

## Author Contributions

**Conceptualization:** Borin Keo, Bin Li.

**Data curation:** Borin Keo.

**Formal analysis:** Borin Keo.

**Investigation:** Borin Keo.

**Methodology:** Borin Keo.

**Project administration:** Borin Keo.

**Resources:** Borin Keo.

**Software:** Borin Keo.

**Supervision:** Bin Li.

**Validation:** Borin Keo.

**Visualization:** Borin Keo.

**Writing – original draft:** Borin Keo.

**Writing – review & editing:** Borin Keo, Waqas Younis.

PLOS ONE  Measuring trade costs and analyzing the determinants of trade growth of Cambodia: 1993–2019PLOS ONE  Measuring trade costs and analyzing the determinants of trade growth of Cambodia: 1993–2019